\documentclass[aps, prb, twocolumn,amsmath,amssymb,floatfix, reprint, longbibliography]{revtex4-2}

\usepackage{graphicx}
\usepackage{svg}
\usepackage{times}
\usepackage{dcolumn}
\usepackage{hyperref}
\usepackage{physics}
\usepackage{siunitx}
\hypersetup{colorlinks, allcolors=blue}
\usepackage{lipsum}
\usepackage{comment}
\usepackage[labelfont=rm]{caption}
\usepackage{subcaption}

\newcommand{\kk}{{\mathbf{k}}}
\newcommand{\mm}{{\mathbf{m}}}
\newcommand{\rr}{{\mathbf{r}}}

\newcommand{\bq}{{\mathbf{q}}}
\newcommand{\e}[1]{\mathrm{e}^{#1}}
\newcommand{\bsigma}{{\boldsymbol{\sigma}}}
\newcommand{\kF}{{k_{\text{F}}}}
\newcommand{\kFs}[1]{{k_{\text{F}#1}}}

\DeclareMathOperator{\sgn}{sgn}

\begin{document}
\title{Dynamically generated spin-interactions and nutational spin inertia in normal metal-ferromagnet heterostructures}
\author{Christian Svingen Johnsen}
\affiliation{\mbox{Center for Quantum Spintronics, Department of Physics, Norwegian University of Science and Technology, NO-7491 Trondheim, Norway}}
\author{Asle Sudb{\o}}
\email[Corresponding author: ]{asle.sudbo@ntnu.no}
\affiliation{\mbox{Center for Quantum Spintronics, Department of Physics, Norwegian University of Science and Technology, NO-7491 Trondheim, Norway}}

\begin{abstract}
We consider the spin dynamics of a normal metal-ferromagnet heterostructure, with emphasis on spin-nutation terms arising from a dynamical Ruderman-Kittel-Kasuya-Yosida (RKKY) interaction. We find that the spin-nutation term is anisotropic in spin space due to the broken time-reversal symmetry of the ferromagnet. This contrasts with what one obtains in the paramagnetic state, where the nutation term is isotropic in spin space. We compute the effects this has on the magnetization dynamics derived from a Landau-Lifshitz-Gilbert equation. In particular, due to broken time-reversal symmetry, we predict a {\it third} ferromagnetic resonance due to nutational  spin dynamics. This resonance frequency is tunable by applying an external magnetic field. We propose this as a strong indicator for the existence of nutation in spin systems. 
\end{abstract}

\maketitle
\section{Introduction}
Van der Waals (vdW) magnets, with their ultra-thin, layer-by-layer structures, open up new possibilities in quantum materials, permitting manipulation of magnetism at the atomic scale and exploring phases of matter that could impact spintronics and nanotechnology in a major way.
These materials are therefore interesting both in terms of fundamental physics as well as in an applications perspective \cite{Burch2018}. An increasing number of such essentially two-dimensional magnets have recently been discovered and exfoliated down to monolayer thickness \cite{Gong2017,Deng2018,Bonilla2018}. In addition to bringing two-dimensional spin systems such as the 2D Heisenberg model and topological phase transitions \cite{Kosterlitz1973,Kosterlitz1974} closer to being realized experimentally, they are expected to provide the field of spintronics \cite{zutic2004} with tunable magnetism \cite{Hendriks2024,Tang2022} which can easily be incorporated in heterostructures with other materials \cite{Geim2013} for potential energy-efficient devices. Furthermore, vdW magnets may bring the past few decades of research on ultrafast magnetization dynamics and inertial effects \cite{Mondal2023} into contact with the field of spintronics, where operating devices at ultrashort timescales is desirable for faster computation.

Due to the proximity effect, layered heterostructures may possess emergent physics not present in each individual layer \cite{Hellman2017,Huang2020}, or one part of the structure can be manipulated via another, easier-to-control material. For instance, some aspects of the spin dynamics of structures like the one considered in this paper, a magnet proximity coupled to a normal metal, were studied a few years ago \cite{Cheng2017,Erlandsen2022}. It was shown that the heterostructure will host an induced \emph{magnon} spin current when a charge current is driven through the metallic part of the structure. An advantage of transporting spin with these currents instead of with electron spin currents \cite{Kajiwara2010} is that the latter causes unnecessary Joule heating because of the motion of the conduction electrons. 

The two abovementioned studies considered spin transport in a regime where the conduction electron dynamics take place at a much shorter timescale than the localized-spin dynamics. In the context of \emph{ultrafast} magnetization dynamics in a metal-ferromagnet heterostructure, however, the magnetization changes over time scales comparable to those of the conduction electrons, meaning interactions with electrons can no longer be considered to be instantaneous, leading to retardation effects or \emph{inertia effects} \cite{Ciornei2011,Fahnle2011,Bhattacharjee2012,Kikuchi2015,Quarenta2024}. These effects have been shown to emerge not only when coupling to a particle bath, but also in pure spin systems as a relativistic correction to a Dirac particle's magnetization in the presence of an electromagnetic field \cite{Mondal2017,Mondal2018}. Another, much earlier derivation of magnetic inertia effects was based on a phenomenological coupling between the magnetization and lattice distortions, giving rise to a memory effect where the magnetization at time $t$ depended on the magnetization at time $t^\prime < t$ \cite{Suhl1998}.

This memory effect appears as an integral over previous magnetic states in the Landau-Lifshitz-Gilbert (LLG) equation \cite{Landau1935,Gilbert2004},
\begin{align}
    \dot{\vb{m}}(t) = \vb{m}(t) \cross \left (  \vb{H}_\text{eff} + \int_{-\infty}^t \dd{t^\prime} \boldsymbol{\chi}(t-t^\prime) \cdot \vb{m}(t^\prime) \right),
\end{align}
where the matrix $\boldsymbol{\chi}$ is derived from whichever effective theory of external degrees of freedom one is considering \cite{Suhl1998,Bhattacharjee2012}. Here, $\vb{m}(t^\prime)$ is usually Taylor expanded in time, yielding a series of time derivatives of $\vb{m}(t)$. The $\dot{\vb{m}}(t)$ term is what gives rise to Gilbert damping towards the effective field $\vb{H}_\text{eff}$. The next correction, $\ddot{\vb{m}}(t)$, is usually called the inertia term \cite{Ciornei2011, Bhattacharjee2012} because of its resemblance to the acceleration appearing in Newton's second law of motion. We will, however, refer to it as the \emph{nutation term} to distinguish it from the known second-time-derivative term that arises in the context of antiferromagnets \cite{Kimel2009}. This choice is motivated by the fact that in ferromagnets (FMs), the nutation term is predicted to lead to nutational motion of the magnetization as it precesses.

The nutation term also leads to a second ferromagnetic resonance (FMR) peak in the THz range for ferromagnets, as opposed to the usual ferromagnetic resonance peak, which is in the GHz range \cite{Olive2012,Neeraj2020}. Excitation of \emph{nutation spin waves} with THz frequency  -- a collective nutational excitation superposed on the more familiar precessional spin waves -- has since been studied theoretically \cite{Lomonosov2021,Cherkasskii2021,Titov2022,Mondal2022,Mondal2023}. Furthermore, the measurement of a half-terrahertz resonance peak in a thin-film ferromagnet was recently reported and attributed to the presence of inertial spin dynamics \cite{Neeraj2020}. Signals of inertial effects in cobolt thin films have also been reported \cite{Unikandanunni2022}.

In this paper we will study the spin dynamics of a thin normal metal (NM) coupled to a two-dimensional ferromagnetic insulator (FMI) -- or the dynamics of a two-dimensional metallic ferromagnet; the distinction is transparent to the theory presented here -- to ascertain what effect conduction electrons being coupled through an sd-like coupling to localized spins has. In essence, it will lead to an effective spin-spin coupling $\boldsymbol{\chi}(\vb{r} - \vb{r}^\prime, t-t^\prime)$ mediated by particle-hole excitations in the NM. This indirect exchange interaction between localized spins mediated by conduction electrons is known as the Ruderman–Kittel–Kasuya–Yosida (RKKY) interaction \cite{Ruderman1954,Kasuya1956,Yosida1957}. The static, or zero-frequency, part of the induced RKKY interaction in $d$ dimensions for a metal was computed in Ref. \cite{Aristov1997}. We extend these results to the full dynamic RKKY interaction of a state with nonzero magnetization.

The nutation term of a metallic ferromagnet has been investigated in three spatial dimensions using a spin gauge field transformation technique, finding a spin-isotropic nutation term for a nonmagnetized state \cite{Kikuchi2015}. Here, we present a more straightforward field-theory calculation, now in two dimensions and including spin ordering directly in the derivation. This serves as a model of vdW magnets, which typically have some degree of spin anisotropy, thus circumventing the Mermin-Wagner theorem \cite{Mermin1966} and allowing for spin ordering even in two dimensions. From the dynamic RKKY interaction, we obtain the nutation term, which will turn out to be significantly spin anisotropic due to the broken symmetry. 

\section{Generated Spin Interactions}
We consider a FMI with a Hamiltonian $H_{\text{FM}}$ containing easy-axis anisotropy along the $z$ axis. The FMI is coupled to a system of gapless fermions with a Hamiltonian given by
\begin{align}
    H_{\text{NM}} &= -\int \dd[2]{r} c^\dagger(\rr) \left ( \frac{\hbar^2}{2m}\boldsymbol{\grad}^2 + \mu \right) c(\rr).
\end{align}
The precise form of $H_\text{FM}$ is not of great importance here, as we will focus on the \emph{additional} spin-spin interactions are generated by the coupling to the gapless fermions. We consider spin fluctuations $\vb{S}(\rr)$ around the mean-field value $\tilde{m}_0 \hat{\vb{z}}$, such that the interaction Hamiltonian becomes
\begin{align}
    H_\text{int} &= -2\bar{J} \int \dd[2]{r} c^\dagger(\rr) \boldsymbol{\sigma} c(\rr) \cdot \bigl[ \vb{S}(\rr) + \tilde{m}_0 \hat{\vb{z}} \bigr],
\end{align}
where $\bar{J}$ is the strength of the interfacial exchange coupling. 
To compute the effective field theory of the spins, we partially compute the partition function,
\begin{equation}
    Z = \int \mathcal{D} \vb{S} \mathcal{D} \psi \mathcal{D} \psi^\dagger \e{-S_\mathrm{FM}[\vb{S}] - S_\mathrm{NM} [\psi, \psi^\dagger] - S_\mathrm{int} [\psi, \psi^\dagger, \vb{S}]},
\end{equation}
by tracing out the fermions $\psi$ in the NM to obtain an effective theory for the spins in the FM by identifying $S_\mathrm{eff}[\vb{S}]$ in
\begin{subequations}
\begin{align}
    Z &= \int \mathcal{D} \vb{S} \e{-S_\mathrm{eff}[\vb{S}]} \\
    &= \int \mathcal{D} \vb{S} \e{-S_\mathrm{FM}[\vb{S}]} \int \mathcal{D} \psi \mathcal{D} \psi^\dagger \e{-\psi^\dagger\left(G^{-1} + B\right)\psi}.
\end{align}
\end{subequations}
Some of the details of this calculation are shown in Appendix \ref{sec:integrate_out_fermions}. The Green's function $G$ for the fermions and the spin-dependent matrix $B$ are given in the Appendix \ref{sec:integrate_out_fermions}. The standard method of deriving a nutation term is to derive an equation of motion for the spins from $S_\mathrm{eff}[\vb{S}]$ using the Euler-Lagrange equations. It describes the magnetization dynamics, taking the form of a generalized Landau-Lifshitz-Gilbert (LLG) equation \cite{Landau1935,Gilbert2004}. To second order in spin-spin interactions, the effective action is
\begin{equation} \label{eq:Seff_TrABAB}
\begin{aligned}
    S_\text{eff}[\vb{S}] &\approx  S_{\text{FM}}[\vb{S}] - \Tr \ln G^{-1} + \Delta S_\text{eff}^{(1)}[\vb{S}]\\
    &\phantom{QQ} +\Delta S_\text{eff}^{(2)}[\vb{S}].
\end{aligned}
\end{equation}
The first two terms are the non-interacting parts coming from the FM and NM, respectively. The second term,
\begin{equation}
    \Delta S_\text{eff}^{(1)}[\vb{S}] = -\Tr \left (G B\right),
\end{equation}
is a spatially uniform correction due to the magnetization $\tilde{m}_0$, affecting only $S_z$. Here, we focus on the induced spin-spin interaction term,
\begin{widetext}
\begin{subequations}
    \begin{align}
        S_\text{eff}^{(2)}[\vb{S}] &= \frac{1}{2} \Tr \left ( G B G B\right)\\ \label{eq:S_eff_after_spin_trace}
        &=\frac{4\bar{J}^2}{\beta^{2}} \int \frac{\dd[2]{k_1}}{(2\pi)^2}\int \frac{\dd[2]{k_2}}{(2\pi)^2}\sum_{n_1,n_2} \frac{S_\alpha(k_1-k_2) S_\alpha(k_2-k_1)}{D(k_1) D(k_2)} \left[d_0(k_1) d_0(k_2) + m_0^2(2\delta_{\alpha z} - 1)\right],
    \end{align}
\end{subequations}
\end{widetext}
where we define the inverse temperature $\beta = 1 / k_\mathrm{B} T$ and
\begin{subequations}
    \begin{align}
    D(k_i) &= (i\omega_{n_i} - E_{\kk_i +})(i\omega_{n_i} - E_{\kk_i -}) \\
    E_{\kk\pm} &= \varepsilon_\kk -\mu  \pm m_0\\
    \varepsilon_{\kk} &= \frac{\hbar^2}{2m}\kk^2\\
    d_0(k_i) &= -i\omega_{n_i} - \mu + \varepsilon_{\kk_i},
\end{align}
\end{subequations}
using the shorthand $k_i = (i \omega_{n_i}, \kk_i)$.
At this stage, the effect of the magnetization $\tilde{m}_0$ in the $z$ direction, now rescaled as $m_0 = -2\bar{J}m_0$, appears as a spin-anisotropic $m_0^2$ term in the effective action. 

\subsection{Real-Time Action}
In the above term of the action, we perform one of the Matsubara sums, such that what is left is a sum over the difference $\omega = \omega_{n_1} - \omega_{n_2}$. Working towards an equation of motion for the spin dynamics, we now take the zero-temperature limit $\beta \to \infty$ and Wick rotate the spin fields to real time by an analytic continuation replacing $i\omega \to \omega + i\delta$, where $\delta$ is an infinitesimal quantity to avoid integrating over poles. More details about this procedure can be found in Appendix \ref{sec:to_real_time}. The resulting action is read off of $\int \mathcal{D}\vb{S}\,\exp (iS_{\text{eff}}^{\text{(RT)}})$. The second-order contribution to the real-time action is now given by
\begin{align}\label{eq:action_real_time_q_omega}
    &\Delta S_\text{eff}^{(\text{2 RT})}[\vb{S}] = \frac{-4\bar{J}^2}{2\pi }\int \frac{\dd[2]{q}}{(2\pi)^2}\int\dd{\omega} \\\nonumber
    &\phantom{QQQQQQQ}\times S_\alpha(\bq, \omega + i\delta)  S_\alpha(-\bq, -\omega -i\delta) \chi_\alpha(\bq, \omega),
\end{align}
where the quantity $\chi_\alpha(\bq, \omega)$ is the Fourier-transformed version of the, in general, nonlocal interaction between spins at different locations and times. Its nonlocality reflects the fact that these spin-spin interactions are mediated by fermions in the NM, which propagate at \emph{finite} speed. The interaction for general $\alpha$ is
\begin{equation}
\label{eq:chi_alpha_start}
    \begin{aligned}
        &\chi_\alpha(\bq, \omega) = \sum_{s_1, s_2 =\pm} \frac{1+s_1 s_2(2\delta_{\alpha z} -1)}{4} \\
        &\phantom{Q}\times \int \frac{\dd[2]{k}}{(2\pi)^2} \frac{f(E_{\kk+\bq, s_1}) - f(E_{\kk, s_2})}{\omega - \bigl [\varepsilon_{\kk+\bq} - \varepsilon_{\kk} + (s_1-s_2)m_0 \bigr] + i\delta},
    \end{aligned}
\end{equation}
where the momentum integration can be performed analytically by using the residue theorem on its angular part. Once again, the procedure is outlined in Appendix \ref{sec:to_real_time}.

\subsection{A Position-Dependent Nutation Term?}
\label{sec:omega_to_time}
Before giving the closed-form expression for $\chi_\alpha(\bq, \omega)$, we will introduce some notation in preparation for the coming sections. Eq.\ \eqref{eq:chi_alpha_start} contains the \emph{nutation term} $I_\alpha^{(2)}(\bq)$, as well as a dissipative term $I_\alpha^{(1)}(\bq)$ and a static RKKY term $I_\alpha^{(0)}(\bq)$. These can be found by transforming the spin fields to the time domain and Taylor expanding them in one of the time coordinates. Here, however, it is more convenient to expand in $\omega$ first and transform to the time domain last. We will then be able to read off the terms of interest from
\begin{align}\label{eq:chi_expansion_prototype}
    \chi_\alpha (\bq, \omega) \approx I_\alpha^{(0)}(\bq) + I_\alpha^{(1)}(\bq)\omega + I_\alpha^{(2)}(\bq) \omega^2.
\end{align}
We should note here that the above equation defines $I_\alpha^{(n)}(\bq)$, but the smallness parameter we actually expand in is $\omega / m_0$, as will become clear from Eq.\ \eqref{eq:full_chi}. The above-claimed interpretation of the $\omega$ coefficients becomes clearer when considering Eq.\ \eqref{eq:action_real_time_q_omega} in the time domain since each power of $\omega$ can be written as a time derivative:
\begin{align}\nonumber
    &\Delta S_\text{eff}^{(\text{2 RT})}[\vb{S}] = \frac{-4\bar{J}^2}{(2\pi)^{2} }\int \frac{\dd[2]{q}}{(2\pi)^2}\int\dd{t_1}\dd{t_2}S_\alpha(\bq, t_1)   \\
    & \times S_\alpha(-\bq, t_2)\e{-\delta(t_1-t_2)} \int \dd{\omega} \left [I_\alpha^{(0)}(\bq) \vphantom{\left(\frac{\partial_{t_2}}{-i}\right)^2}\right. \\\nonumber 
    &\phantom{QQQQQq} \left.+ I_\alpha^{(1)}(\bq)\frac{\partial_{t_2}}{-i} + I_\alpha^{(2)}(\bq) \left(\frac{\partial_{t_2}}{-i}\right)^2\right] \e{i\omega(t_1-t_2)}.
\end{align}
Performing the frequency integral makes the action local in time except for time derivatives on $S_\alpha(-\bq, t_2)$, which come about after integrating by parts and discarding the surface terms. The resulting action is
\begin{align}\nonumber
    &\Delta S_\text{eff}^{(\text{2 RT})}[\vb{S}] = \frac{-4\bar{J}^2}{2\pi}\int \frac{\dd[2]{q}}{(2\pi)^2}\int\dd{t}S_\alpha(\bq, t)  \\
    & \times\left [ I_\alpha^{(0)}(\bq)  S_\alpha(-\bq, t) + iI_\alpha^{(1)}(\bq) \dot{S}_\alpha(-\bq, t)\right. \\\nonumber
    &\phantom{QQ} \left. - I_\alpha^{(2)}(\bq) \ddot{ S}_\alpha(-\bq, t) \right].
\end{align}

In the literature, one typically takes the time derivatives of the spins to be position independent to create a spatially local nutation term \cite{Ciornei2011,Bhattacharjee2012,Quarenta2024,Kikuchi2015}. These types of nutation terms are the ones which have been shown to lead to nutational motion of the spin superposed on the regular damped precessional motion.
If, instead, one keeps the $I_\alpha^{(n)}(\bq)$ momentum dependent, the nutation term becomes much more complicated, now containing an integral over all spins near spin $S_\alpha(\rr, t)$ because 
\begin{align}\nonumber
    &\Delta S_\text{eff}^{(\text{2 RT})}[\vb{S}] = \frac{- 4\bar{J}^2}{2\pi}\int \dd[2]{r}\int \dd{t}S_\alpha(\vb{r}, t) \int\dd[2]{(\Delta r)} \\ \nonumber
    &\times \left [ I^{(0)}_\alpha(\Delta \vb{r}) S_\alpha(\vb{r} - \Delta \vb{r}, t)  + iI_\alpha^{(1)}(\Delta \vb{r}) \dot{S}_\alpha(\vb{r} - \Delta \vb{r}, t) \right. \\\label{eq:general_inertial_action}
    &\phantom{QQ} \left.- I_\alpha^{(2)}(\Delta \vb{r}) \ddot{S}_\alpha(\vb{r} - \Delta \vb{r}, t)\right].
\end{align}
An equation of motion for the spins, the LLG equation, can be derived from this action by rewriting the Euler-Lagrange equations. In general, these can be expressed as $\tilde{\boldsymbol{L}} = 0$, where the vector $\tilde{\boldsymbol{L}}$ is a sum of various derivatives of the effective Lagrangian derived from the effective action $S_\text{eff}[\vb{S}]$. Since we have not integrated by parts to remove the double time derivative $\ddot{S}_\alpha$, we need a second-derivative term in $\tilde{\boldsymbol{L}}$ as well. The term in $\tilde{\boldsymbol{L}}$  stemming from the spin-spin interaction term in the action is given by
\begin{equation}\label{eq:EL_2}
\begin{aligned}
    L_i^{(\text{2})}(\rr, t) &=  \pdv{\mathcal{L}_\text{eff}^{(2)}(\rr, t)}{m_i(\rr, t)} - \dv{}{x_\mu}\pdv{\mathcal{L}_\text{eff}^{(2)}(\rr, t)}{\left(\pdv{m_i(\rr, t)}{x_\mu}\right)} \\
    &\phantom{=} + \dv[2]{}{x_\mu}\pdv{\mathcal{L}^{(2)}_\text{eff}(\rr, t)}{\left(\pdv[2]{m_i(\rr, t)}{x_\mu}\right)},
\end{aligned}
\end{equation}
where $i=x,y,z$ and $x_\mu=(x, y, t)$, and where summation over $\mu$ is implied, while $i$ is now fixed. The Lagrangian above is defined through $\Delta S_\text{eff}^{(\text{2 RT})}[\vb{m}] = \int \dd[2]{r}\dd{t}\mathcal{L}_\text{eff}^{(2)}(\rr, t)$, and the spins $S_\alpha(\rr, t)$ are now reinterpreted as a magnetization $m_i(\rr, t)$. It can be shown that this vector $L_i^{(2)}(\rr,t)$ contributes additively to the vector $L_i(\rr,t)$ which enters on the right-hand side of the LLG equation,
\begin{align}\label{eq:LLG_general_L}
    \dot{\mm}(\rr, t) = \mm(\rr, t) \cross \vb{L}(\rr, t).
\end{align}
The main idea behind the derivation of Eq.\ \eqref{eq:LLG_general_L} is to take into account the Berry phase term $\vb{b}\cdot \dot{\mm}$ in $\mathcal{L}_\text{FMI}$, where the Berry connection $\vb{b}$ satisfies $\partial_\mm \cross \vb{b} = -\mm/\mm^2$. In the present case, the additional terms in the LLG equation are given by
\begin{widetext}
\begin{equation}
    \begin{aligned}
    &L^{(2)}_i(\rr, t) = \frac{- 4\bar{J}^2}{2\pi} \int \dd[2]{(\Delta r)} \left [ I^{(0)}_i(\Delta \vb{r}) m_i(\vb{r} - \Delta \vb{r}, t)  + iI_i^{(1)}(\Delta \vb{r}) \dot{m}_i(\vb{r} - \Delta \vb{r}, t)  - I_i^{(2)}(\Delta \vb{r}) \ddot{m}_i(\vb{r}- \Delta \vb{r}, t)\right] \\ 
    &\phantom{QQQQQQQQQQq} + \frac{-4\bar{J}^2}{2\pi} \left [ I^{(0)}_i(\Delta \rr =0) m_i(\rr,t)  - iI_i^{(1)}(\Delta \rr =0) \dot{m}_i(\rr,t)  - I^{(2)}_i(\Delta \rr = 0)\ddot{m}_i(\rr,t) \right]
\end{aligned}
\end{equation}
\end{widetext}
when the time derivatives of $\mm$ are allowed to vary in space. Allowing such position dependence in $\ddot{m}_i$ and $I_i^{(2)}$ will turn out to be challenging due to momentum-space divergences. To get the commonly-reported LLG equation that we will solve numerically in a later section, we will go back to the term $\Delta S_\text{eff}^{(\text{2 RT})}[\vb{S}]$ in the action instead. Inserting the assumption of spatially independent \emph{spin} dynamics at the action level, before making the LLG equation, will lead to the desired form of the LLG. The assumption amounts to setting $\dot{S}_\alpha(\rr - \Delta \rr, t) \approx \dot{S}_\alpha(\rr, t)$ and $\ddot{S}_\alpha(\rr - \Delta \rr, t) \approx \ddot{S}_\alpha(\rr, t)$, which means the $\Delta r$ integral only acts on $I^{(1)}$ and $I^{(2)}$ in the derivative terms, yielding a delta function on the momentum $\bq$ in $I_\alpha^{(n)}(\Delta \rr) = \int\dd[2]{q} I^{(n)}_\alpha(\bq)\e{i\bq \cdot \Delta \rr}/(2\pi)^2$. Eq.\ \eqref{eq:general_inertial_action} now assumes the simpler form
\begin{align}\nonumber
    &\Delta S_\text{eff}^{(\text{2 RT})}[\vb{S}] = \frac{- 4\bar{J}^2}{2\pi}\int \dd[2]{r}\int \dd{t}S_\alpha(\vb{r}, t) \\ \label{eq:inertial_action_position_independence}
    &\times \left [\int\dd[2]{(\Delta r)} I^{(0)}_\alpha(\Delta \vb{r}) S_\alpha(\vb{r} - \Delta \vb{r}, t)  \right. \\\nonumber
    &\phantom{QQ} \left.+ iI_\alpha^{(1)}(\bq=0) \dot{S}_\alpha(\vb{r}, t) - I_\alpha^{(2)}(\bq = 0) \ddot{S}_\alpha(\vb{r}, t) \vphantom{\int}\right],
\end{align}
meaning
\begin{align}\nonumber
    &L^{(2)}_i(\rr, t) = \frac{- 4\bar{J}^2}{2\pi} \left [ \int \dd[2]{(\Delta r)} I^{(0)}_i(\Delta \vb{r}) m_i(\vb{r} - \Delta \vb{r}, t)  \right. \\\label{eq:LLG_L_spatial_indep}
    &\left. - 2I_i^{(2)}(\bq=0) \ddot{m}_i(\vb{r}, t)+ I^{(0)}_i(\Delta \rr =0) m_i(\rr,t) \vphantom{\int} \right],
\end{align}
where the dissipation term cancels out in the $q=0$ limit taken here. Mathematically, the cancellation is due to the minus sign in Eq.\ \eqref{eq:EL_2}, and physically, we cannot expect that this limit yields dissipation since the dissipative limit would be  $q\ll1$ \emph{and} $\omega / q v_\text{F} \ll 1$, where $v_\mathrm{F}$ is the Fermi velocity \cite{Hertz1976}. From Eq.\ \eqref{eq:LLG_L_spatial_indep}, $\vb{I}$ in the nutation term $\mm \cross (\vb{I}\cdot \ddot{\mm})$ on the right-hand side of Eq.\ \eqref{eq:LLG_general_L} can be readily read off as
\begin{align}
    \vb{I} = \frac{4\bar{J}^2}{\pi} \begin{pmatrix}
        I_x^{(2)}(\bq=0) & 0 & 0\\
        0 & I_y^{(2)}(\bq=0) & 0\\
        0 & 0 & I_z^{(2)}(\bq=0)
    \end{pmatrix}.
\end{align}
Now that we have established a connection to the language of previous studies, we move on to our results.  

In this paper, we are primarily concerned with the nutation term $I^{(2)}_\alpha$ and the static RKKY term $I^{(0)}_\alpha$, i.e., the terms with even powers of $\omega$ in Eq.\ \eqref{eq:chi_expansion_prototype}. This means it suffices to consider the real part of $\chi_\alpha(\bq, \omega)$ because the imaginary part is odd in $\omega$, whereas the real part is even. These (anti)symmetries of $\chi_\alpha(\bq, \omega)$ can be seen from Eq.\ \eqref{eq:chi_alpha_start} by relabeling or shifting the dummy variables $s_1, s_2,$ and $k$. The exact form of the real part of the spin-spin interaction reads
\begin{widetext}
\begin{subequations}
\label{eq:full_chi}
\begin{align}
    &\Re \chi_{\alpha \neq z}(\bq, \omega) = \frac{2m}{\hbar^2}\frac{1}{8\pi} \sum_{s=\pm} \left [1  - \sgn(\nu_+(s)) \theta\left ( \abs{\frac{\nu_+(s)}{2}} - \sqrt{\frac{\mu}{m_0}}\frac{\kFs{s}}{\kF} \right) \frac{1}{\bar{q}}\sqrt{\left(\frac{\nu_+(s) }{2} \right)^2 -\frac{\mu}{m_0}\left(\frac{\kFs{s}}{\kF}\right)^2 } \right.\\\nonumber
    &\phantom{QQQQQQQQQQQQQQQQQ}\left. + \sgn(\nu_-(s)) \theta\left ( \abs{\frac{\nu_-(s)}{2}} - \sqrt{\frac{\mu}{m_0}}\frac{\kFs{s}}{\kF} \right) \frac{1}{\bar{q}}\sqrt{\left(\frac{\nu_-(s) }{2} \right)^2 -\frac{\mu}{m_0}\left(\frac{\kFs{s}}{\kF}\right)^2 } \right],\\
    &\Re \chi_{z}(\bq, \omega) = \frac{2m}{\hbar^2} \frac{1}{8 \pi} \sum_{s=\pm} \left [ 1 - \sgn \left( \nu + \bar{q}^2 \right) \theta\left ( \abs{\frac{\nu+\bar{q}^2}{2\bar{q}}} - \sqrt{\frac{\mu}{m_0}}\frac{\kFs{s}}{\kF} \right) \frac{1}{\bar{q}}\sqrt{\left(\frac{\nu + \bar{q}^2}{2\bar{q}} \right)^2 -\frac{\mu}{m_0}\left(\frac{\kFs{s}}{\kF}\right)^2 } \right.\\\nonumber
    &\phantom{QQQQQQQQQQQQQQQQ} \left.+ \sgn \left( \nu - \bar{q}^2 \right) \theta\left ( \abs{\frac{\nu-\bar{q}^2}{2\bar{q}}} - \sqrt{\frac{\mu}{m_0}}\frac{\kFs{s}}{\kF} \right) \frac{1}{\bar{q}}\sqrt{\left(\frac{\nu - \bar{q}^2}{2\bar{q}} \right)^2 -\frac{\mu}{m_0}\left(\frac{\kFs{s}}{\kF}\right)^2 } \right ],
\end{align}
\end{subequations}
\end{widetext}
where the step function $\theta(x)$ is 0 for $x < 0$ and 1 for $x>0$.  Here we have introduced rescaled momenta,
\begin{align}
    \bar{q} = \sqrt{\frac{\mu}{m_0}} \frac{q}{\kF},
\end{align}
and energies,
\begin{align}  
    \nu = \frac{\omega}{m_0},
\end{align}
in addition to the shorthand
\begin{align}
    \nu_\pm(s) = \frac{\nu \pm \left ( \bar{q}^2 - 2s \right)}{\bar{q}}.
\end{align}
The above expressions for $\Re\chi_\alpha$ will later be expanded in $\omega/m_0$ to get the coefficients in Eq.\ \eqref{eq:chi_expansion_prototype}. We will eventually choose to express our results using the dimensionless momentum $\tilde{q} = q / \kF$.

\section{Static RKKY}
\begin{figure*}
    \centering
    \begin{subfigure}[b]{0.49\textwidth}
        \centering
        \includegraphics[width=\linewidth]{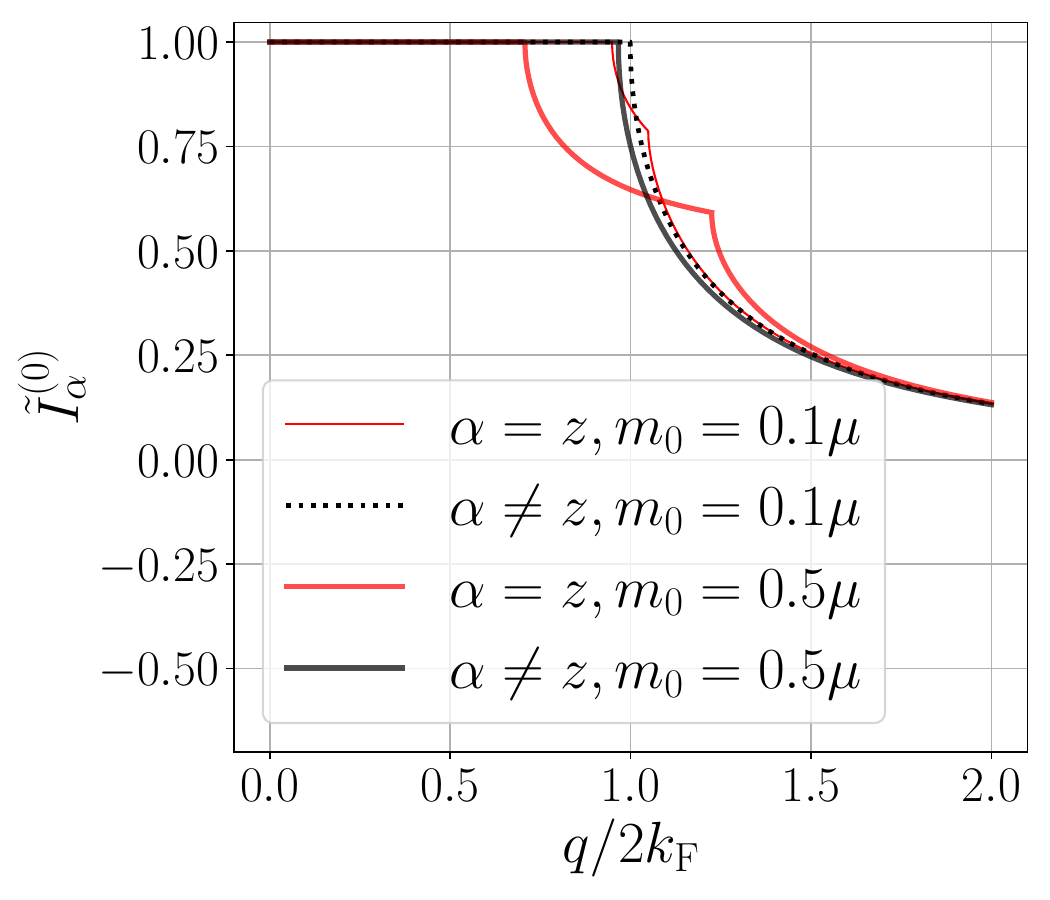}
    \end{subfigure}
    \centering
    \begin{subfigure}[b]{0.49\textwidth}
        \centering
        \includegraphics[width=\linewidth]{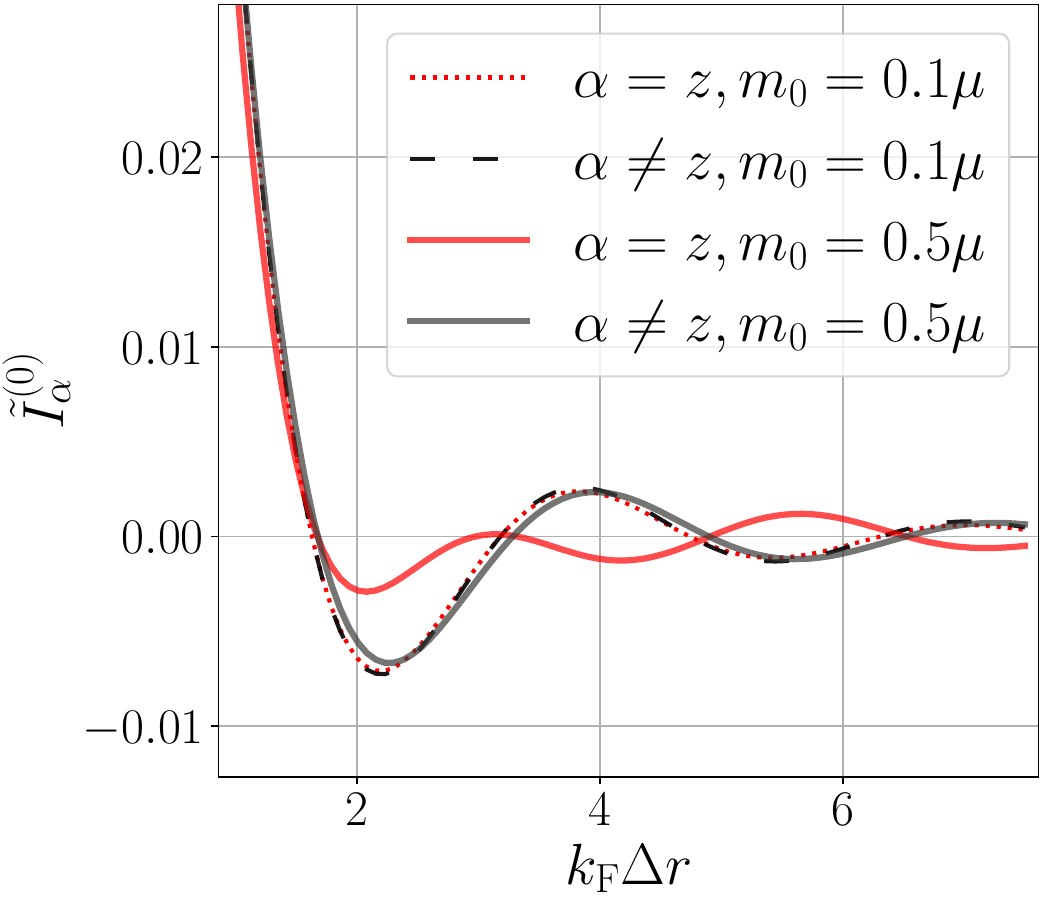}
    \end{subfigure}
    \caption{\small \emph{Left}: Normalized static RKKY term $I^{(0)}_\alpha(q)$ in momentum space. \emph{Right}: Real-space version of the left panel. Both variants display a larger spin anisotropy for larger magnetization $m_0$. The length scales here are essentially set by the ratio $m_0/\mu$, where $\mu$ is the chemical potential.}
    \label{fig:I0_of_rq}
\end{figure*}

The real part of the Lindhard function, or dynamic RKKY interaction, $\Re \chi_\alpha(\bq, \omega)$ has a $\bq, \omega$ dependence that is not easily visualized in a three-dimensional plot, so we will restrict ourselves to the limit $\omega/m_0 \ll 1$ and calculate $I^{(0)}_\alpha(\bq)$ and $I^{(2)}_\alpha(\bq)$ as defined in Eq.\ \eqref{eq:chi_expansion_prototype}. We begin with the regular, static RKKY interaction $I^{(0)}_\alpha(\bq)$ as it can be compared to previous works which considered zero net magnetization $m_0$ \cite{Aristov1997}. We have $I^{(0)}_\alpha(\bq) = \Re \chi_\alpha(\bq, 0)$, so
\begin{align}  \nonumber
    I_z^{(0)}(\bq) & =  \frac{2m}{\hbar^2} \frac{1}{4\pi} \left [ 1 - \frac{\theta(q-2\kF_+)}{2 \Tilde{q}} \sqrt{\Tilde{q}^2 - \left(\frac{2\kFs{+}}{\kF}\right)^2} \right.\\\label{I0z}
    &\phantom{QQQqq}  \left.- \frac{\theta(q-2\kF_-)}{2 \Tilde{q}} \sqrt{\Tilde{q}^2 - \left(\frac{2\kFs{-}}{\kF}\right)^2} \right],
\end{align}
and
\begin{align}
\label{I0xysimple}
    I_{\alpha\neq z}^{(0)}(\bq) &= \frac{2m}{\hbar^2} \frac{1}{4 \pi} \Biggl [ 1 - 
    \frac{\theta(\tilde{q}-\tilde{q}_+)}{\tilde{q}^2}
    \sqrt{\left(\tilde{q}^2-\tilde{q}_-^2\right)\left(\tilde{q}^2-\tilde{q}_+^2\right)} \Biggr ],
\end{align}
where the Fermi momentum $\kF = \sqrt{2m \mu}/\hbar$, $\tilde{q} = q/\kF$, and
\begin{subequations}\label{eq:q_tilde_kF_def}
    \begin{align}
    \tilde{q}_{\pm}^2 &= 2 \left ( 1 \pm  \sqrt{1-\frac{m_0^2}{\mu^2}} \right)\\
    \kF_\pm &= \kF \sqrt{1 \mp \frac{m_0}{\mu}}.
\end{align}
\end{subequations}
The $\alpha = x, y$ terms are equal, but owing to the spin-space anisotropy from the magnetization in the $z$ direction, $I_z^{(0)}(\bq)$ differs somewhat from the other two terms. This is demonstrated in Figure \ref{fig:I0_of_rq}, where these quantities are shown for two values of the magnetization $m_0$. In the figure, the prefactor $2m / 4\pi  \hbar^2$ is left out. 
\subsection{Real-space Static RKKY}
In Figure \ref{fig:I0_of_rq}, the real-space dependence of $I^{(0)}_\alpha$ is shown after numerically integrating over $q$ in
\begin{subequations}
    \begin{align}
        I_\alpha^{(0)}(\Delta \vb{r}) &= \int_0^\infty \frac{\dd{q}}{2\pi} q J_0(q \Delta r) I_\alpha^{(0)}(q),
    \end{align}
\end{subequations}
where the zeroth-order Bessel function of the first kind $J_0(q \Delta r)$ appears because of the angular integration of $\bq$. Again, the result is independent of the direction of $\Delta \vb{r}$. For $\alpha = z,$ it is actually possible to get an exact analytic expression for $I^{(0)}_z(\Delta \rr)$ by extending the results of Ref. \cite{Aristov1997} to finite $m_0.$

The starting point for this calculation is Eq.\ \eqref{eq:S_eff_after_spin_trace}, which in real space takes the form
\begin{align}\nonumber
    \Delta S_\text{eff}^{(2)} &= \frac{4 \bar{J}^2}{\beta}\iint \dd[2]{r_1}\dd[2]{r_2} \sum_{i\omega_\nu} S_\alpha(\rr_1, i\omega_\nu)S_\alpha(\rr_1, -i\omega_\nu)  \\
    &\phantom{QQQQQQQQQQQQQ}\times \chi_\alpha(\rr_1 - \rr_2, i\omega_\nu),
\end{align}
where $\chi_\alpha(\rr_1 - \rr_2, i\omega_\nu)$ is the Matsubara-frequency version of the real-space interaction, and its real-frequency counterpart is $\chi_\alpha(\rr_1 - \rr_2, \omega)$. At zero temperature and zero frequency, these coincide. As outlined in Appendix \ref{sec:aristov}, the interaction takes the form
\begin{align}\label{eq:chi_aristov_comp_GG}
    &\chi_\alpha(\Delta \rr, i\omega_\nu= 0) = -\frac{1}{2\beta} \sum_n \sum_{s_1,s_2=\pm} g_\alpha(s_1,s_2) \\\nonumber
    &\phantom{QQQQQQQQQ} \times G_{s_1}(i\omega_n, \Delta \rr)G_{s_2}(i\omega_n, -\Delta \rr),
 \end{align}
 where
 \begin{align}
    G_\pm(i\omega_n, \Delta \rr) &= \int\frac{\dd[2]{k}}{(2\pi)^2} \frac{\e{i\kk\cdot \Delta \rr}}{i\omega_n - E_{\kk\pm}}
 \end{align}
is the direct generalization of Aristov's function $G(i\omega, \vb{R})$ \cite{Aristov1997} to the case $m_0\neq0$. The prefactor $g_\alpha$ is given by $g_\alpha(\pm, \pm) = \delta_{\alpha z}$, $g_\alpha(\pm, \mp) = 1-\delta_{\alpha z}$. We outline some more details of how to modify Aristov's calculations to finite $m_0$ in Appendix \ref{sec:aristov}. The result for $\alpha = z$ is
\begin{widetext}
\begin{align}\label{chi_z-static}
     \chi_z(\Delta \rr, i\omega_\nu= 0)&= - \frac{2m}{\hbar^2} \frac{1}{16\pi}  \biggl  \{ \kF_{+}^2 \bigl [J_0(\kF_+ \Delta r)Y_0(\kF_+ \Delta r) + J_1(\kF_+ \Delta r)Y_1(\kF_+ \Delta r) \bigr] + \\ \nonumber
    &\phantom{QQQQQQq} + \kF_{-}^2 \bigl [ J_0(\kF_- \Delta r)Y_0(\kF_- \Delta r) + J_1(\kF_- \Delta r)Y_1(\kF_- \Delta r)  \bigr] \biggr\}.
\end{align}
\end{widetext}
In addition to the exact expression for $\chi_z(\Delta \rr, i\omega_\nu=0)$, an approximate analytical expression for the $\alpha \neq z$ case can be found by considering the fact that
\begin{align}
    I^{(0)}_{\alpha \neq z}(\bq) &= \frac{2m}{\hbar^2} \frac{1}{4 \pi} \Biggl [ 1 - 
    \frac{\theta(\tilde{q}-\tilde{q}_+)}{\tilde{q}}
    \sqrt{\left(\tilde{q}^2-\tilde{q}_+^2\right)}F(\tilde{q}) \Biggr ],
\end{align}
where the function $F(\tilde{q}) = \sqrt{1 - \tilde{q}_-^2/\tilde{q}^2}$ is close to unity for $\tilde{q} > \tilde{q}_+$ and $m_0 \lesssim 0.2\mu$. Under the approximation $F(\tilde{q}) \approx 1$, then, we obtain 
\begin{align}
    I^{(0)}_{\alpha \neq z}(\bq) &\approx \frac{2m}{\hbar^2} \frac{1}{4 \pi} \Biggl [ 1 - 
    \frac{\theta(\tilde{q}-\tilde{q}_+)}{2\tilde{q}}
    \sqrt{\left(\tilde{q}^2-\tilde{q}_+^2\right)} \\ \nonumber
    &\phantom{QQQQQQQ} - \frac{\theta(\tilde{q}-\tilde{q}_+)}{2\tilde{q}}
    \sqrt{\left(\tilde{q}^2-\tilde{q}_+^2\right)} \Biggr ],
\end{align}
which is analogous to $I^{(0)}_z(\bq)$ with the replacement $2\kFs{\pm}/\kF \to \tilde{q}_+$ in the square roots and the step functions $\theta$, implying that
\begin{align}
     &\chi_{\alpha\neq z}(\Delta \rr, i\omega_\nu= 0) \approx -\frac{2m}{\hbar^2} \frac{\kF^2}{32\pi} \tilde{q}_+^2 \\\nonumber
     &\times\left [J_0\left(\frac{\tilde{q}_+ \Delta r}{2}\right) Y_0 \left(\frac{\tilde{q}_+ \Delta r}{2}\right)  + J_1\left(\frac{\tilde{q}_+ \Delta r}{2}\right)Y_1\left(\frac{\tilde{q}_+ \Delta r}{2}\right) \right].
\end{align}
For $m_0 \lesssim 0.2\mu$, this is an excellent approximation.

We note that in the limit $m_0\to 0$, the graphs in Fig.\ \ref{fig:I0_of_rq} coincide with those corresponding to the expressions in Ref.\ \cite{Aristov1997}, restoring the spin anisotropy. We now move on to considerations of $\chi_\alpha$'s frequency dependence and the nutation term.

\section{Frequency Dependence and The Nutation Term}
\begin{figure}
    \centering
    \includegraphics[width=\linewidth]{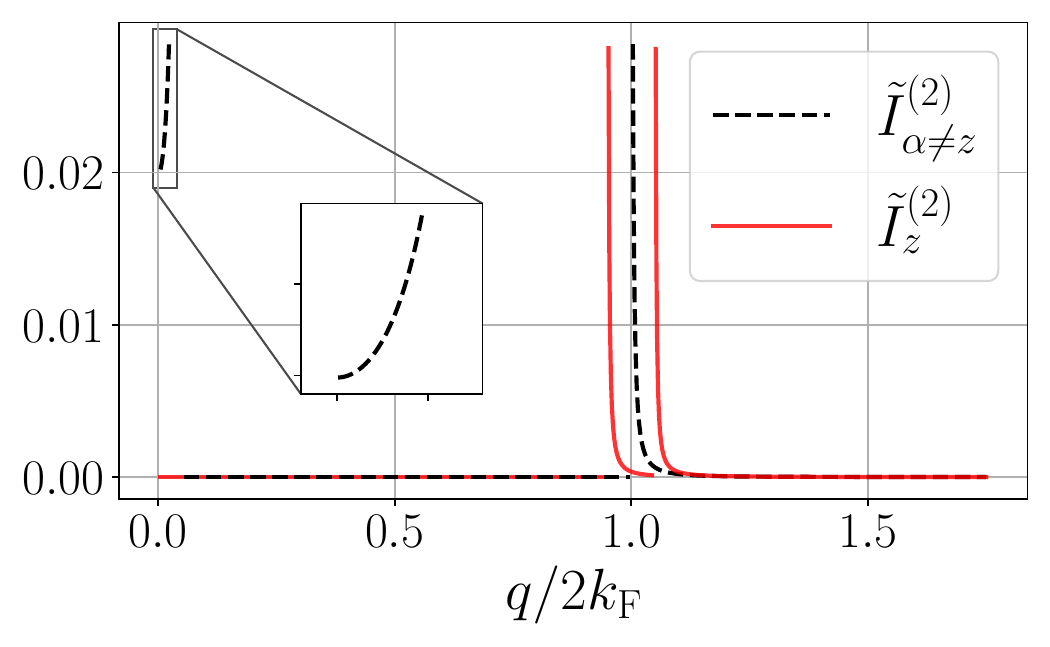}
    \caption{\small Functional form of the momentum dependence of the nutation terms $I^{(2)}_{\alpha}(\bq)$, here scaled by $\hbar^2m_0^2/2m$ to form the dimensionless quantity $\tilde{I}^{(2)}_{\alpha}(\bq)$. They diverge at the momenta $\tilde{q}_\pm$ for $\alpha \neq z$ and at $\kFs{\pm}$ for $\alpha = z$, which only depend on the ratio between the magnetization and the chemical potential, $m_0/\mu=0.1$. We refer to the main text for the details.}
    \label{fig:I2}
\end{figure}
The nutation term stems from $I^{(2)}$ which we obtain by calculating the $\omega^2$ coefficient in Eq.\ \eqref{eq:chi_expansion_prototype}, i.e. Taylor expanding $\Re \chi_\alpha(\bq, \omega)$ to second order in $\omega/m_0$. The result is
\begin{subequations}
\label{eq:I_2_two_singularities}
    \begin{align}
        I_{\alpha\neq z}^{(2)}(\bq) &= \left(\frac{2m}{\hbar^2}\right)^3 \frac{1}{2\pi \kF^4}  \frac{\theta(\tilde{q}_- -\tilde{q}) \frac{m_0}{\mu} + \theta(\tilde{q} - \tilde{q}_+)}{\left[\left(\tilde{q}^2-\tilde{q}^2_+\right) \left(\tilde{q}^2-\tilde{q}^2_-\right)\right]^{3/2}} \\\nonumber
        I_z^{(2)}(\bq)&= \left(\frac{2m}{\hbar^2}\right)^3 \frac{1}{4\pi \kF^4}\frac{1}{\tilde{q}^3} \\
        &\phantom{QQ}\times \left \{ \frac{\left(1- \frac{m_0}{\mu}\right)\theta\left(\tilde{q} - \frac{2\kF_+}{\kF}\right)}{\left[\left(\tilde{q} - \frac{2\kF_+}{\kF}\right) \left(\tilde{q} + \frac{2\kF_+}{\kF}\right) \right]^{3/2}}  \right. \\\nonumber
        &\phantom{QQQ} \left. + \frac{\left(1+ \frac{m_0}{\mu}\right)\theta\left(\tilde{q} - \frac{2\kF_-}{\kF}\right)}{\left[\left(\tilde{q} - \frac{2\kF_-}{\kF}\right) \left(\tilde{q} + \frac{2\kF_-}{\kF}\right) \right]^{3/2}} \right\},
    \end{align}
\end{subequations}
each of which is divergent at two points. Fig.\ \ref{fig:I2} shows these two expressions for $m_0/\mu =0.1$. The length scales are set by the ratio $m_0 / \mu$ in the way shown in Eq.\ \eqref{eq:q_tilde_kF_def}. Considering $\omega \ll m_0$ and $q\ll \kF$, one finds that the contribution to the spatially local nutation term is 
\begin{align} \label{eq:intertia_omega_then_q}
    &I^{(2)}_{\alpha \neq z}(\bq = 0) = \frac{2m}{\hbar^2} \frac{1}{m_0^2} \frac{1}{2^4 \pi},
\end{align}
which is similar to Ref.\ \cite{Kikuchi2015} in that the interfacial exchange interaction strength $\bar{J}$ does not affect the nutation term much. In fact, the cancellation of the $\bar{J}^2$ prefactor in e.g., Eq.\ \eqref{eq:inertial_action_position_independence} and in the above $m_0^2=(-2\bar{J}\tilde{m}_0)^2$ is exact, which is due to the system being two dimensional. Since we have considered finite magnetization $m_0$, there is -- like in the static RKKY case -- a spin-space anisotropy. The $\alpha = z$ term is simply
\begin{align}
    I^{(2)}(\tilde{q}=0) = 0,
\end{align}
a difference which affects the spin dynamics in a few ways.

\section{Landau-Lifshitz-Gilbert Equation with Anisotropic Nutation Term}
In this section, we demonstrate some consequences of the spin anisotropy of the nutation term, including a new ferromagnetic resonance (FMR) signal which should be present when the magnet is ordered and can host nutational dynamics. In the previous section, we calculated the second-order-in-spin contribution to the LLG equation from the conduction electrons up to second order in time derivatives. Here we will simply add in this contribution to the well-known LLG with damping and precession about an external field $\vb{H}$. Thus, as a starting point, we take an LLG equation of the form
\begin{align}
    \dot{\mm} &=  \mm \cross \left [ \gamma\vb{H} + \alpha \dot{\mm} + \left ( \vb{I} \cdot \ddot{\mm} \right)\right]
\end{align}
with $\vb{I} = \text{diag}\left(I, I, \tilde{I}\right)$. We parameterize the anisotropy in the nutation term by the quantity $\tilde{I}$ which in the anisotropic case would be $\tilde{I}=0,$ and $\tilde{I}=I$ in the isotropic case. Employing spherical coordinates,
\begin{align}
    \mm = (\sin \theta \cos\phi, \sin\theta \sin\phi, \cos\theta),
\end{align}
and assuming the field is along the magnetization axis, $\vb{H} = H \hat{\vb{z}}$, the LLG equation can be written in terms of the polar angle $\theta$ and azimuth angle $\phi$ as
\begin{align}\label{eq:llg_theta}
    &\ddot{\theta} \left(\cos^2\theta + \frac{\tilde{I}}{I}\sin^2\theta \right)  = - \frac{\alpha}{I}\dot{\theta} +\frac{1}{I} \dot{\phi} \sin \theta \\\nonumber 
    &\phantom{=} + \dot{\phi}^2 \frac{1}{2}\sin2\theta + \frac{1}{2}\left(1-\frac{\tilde{I}}{I}\right) \dot{\theta}^2 \sin2\theta + \frac{\gamma H}{I} \sin\theta
\end{align}
and
\begin{align}\label{eq:llg_phi}
    \ddot{\phi} \sin\theta  &= -\frac{1}{I} \dot{\theta} - \frac{\alpha}{I} \dot{\phi} \sin\theta - 2 \dot{\theta} \dot{\phi} \cos \theta.
\end{align}
The anisotropy clearly does not affect the $\ddot{\phi}$ equation explicitly, but rather implicitly by modifying the $\ddot{\theta}$ equation. The latter has an extra $\dot{\theta}^2$ term modifying the dynamics. Additionally, the coefficient of $\ddot{\theta}$ -- which for isotropic $\tilde{I}=I$ would be at its maximum value, $1$ -- is \emph{reduced} by the anisotropy, loosely suggesting faster dynamics in the anisotropic case. 

The equations can also be recast in dimensionless form using the rescaled time $\tilde{t}=t/I$, now reading \cite{Olive2015}
\begin{align}\nonumber
    &\theta^{\prime \prime} \left(\cos^2\theta + \frac{\tilde{I}}{I}\sin^2\theta \right)  = - \alpha \theta^\prime + \phi^\prime \sin \theta + \frac{1}{2}(\phi^\prime)^2 \sin 2\theta \\
    &\phantom{QQQ} +\frac{1}{2} \left(1-\frac{\tilde{I}}{I}\right) (\theta^\prime)^2 \sin2\theta+  \gamma HI \sin\theta
\end{align}
and
\begin{align}
    \phi^{\prime\prime} \sin \theta &= -\theta^\prime - \alpha \phi^\prime \sin\theta - 2\theta^\prime \phi^\prime \cos\theta,
\end{align}
where primes represent differentiation with respect to $\tilde{t}$. We will show some numerical solutions of these dimensionless equations in the following paragraphs.

\begin{figure}
    \centering
    \includegraphics[width=0.95\linewidth]{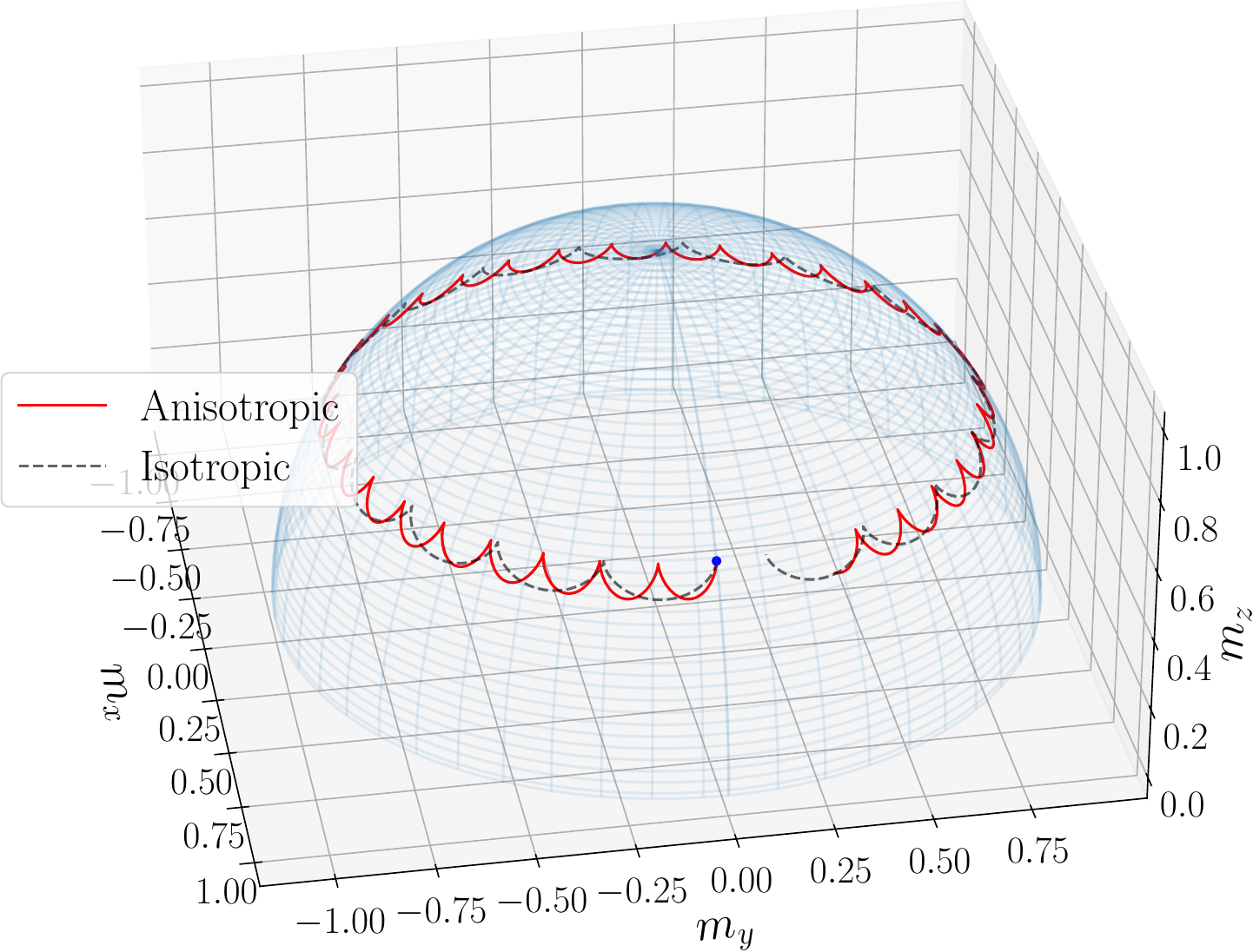}
    \caption{\small Path of the magnetization $\mm(t)$ in a magnetic field pointing in the $z$ direction starting from a motionless state at $t=0$ (blue dot) to $t=38\pi I$ for both an isotropic nutation term and an anisotropic one.}
    \label{fig:faster_dynamics}
\end{figure}

To give a simple example of the faster-dynamics effect mentioned above, we consider $\gamma H I = 0.05$, which corresponds to a nutation timescale $I$ only slightly smaller than the precessional timescale set by the field $H$. This large $I$ is chosen for illustrative purposes; the nutation timescale is expected to be orders of magnitude lower than the precessional one \cite{Olive2012}. We will therefore set the damping $\alpha = 0$, which will serve as an approximation to study dynamics happening at much smaller timescales than the damping time. Fig.\ \ref{fig:faster_dynamics} shows the path of the magnetization $\mm(t)$ from $t=0$, the blue dot, until $t = 38\pi I$, with $\phi(0)=0$, $\theta(0)=\SI{55}{\degree}$, and $\dot{\mm}(0) = 0$ with and without anisotropy. The isotropic case shows precession and nutation with cusps. The nutation there happens at a larger timescale than in the anisotropic case. The latter has more cusps, i.e., faster nutational dynamics. Furthermore, it comes closer to completing a full revolution around the pole, so the precession also seems affected by the anisotropy. This could come about from a third, precession-like resonance in the system, as will become clear shortly.

\begin{figure*}
    \centering
    \includegraphics[width=0.95\linewidth]{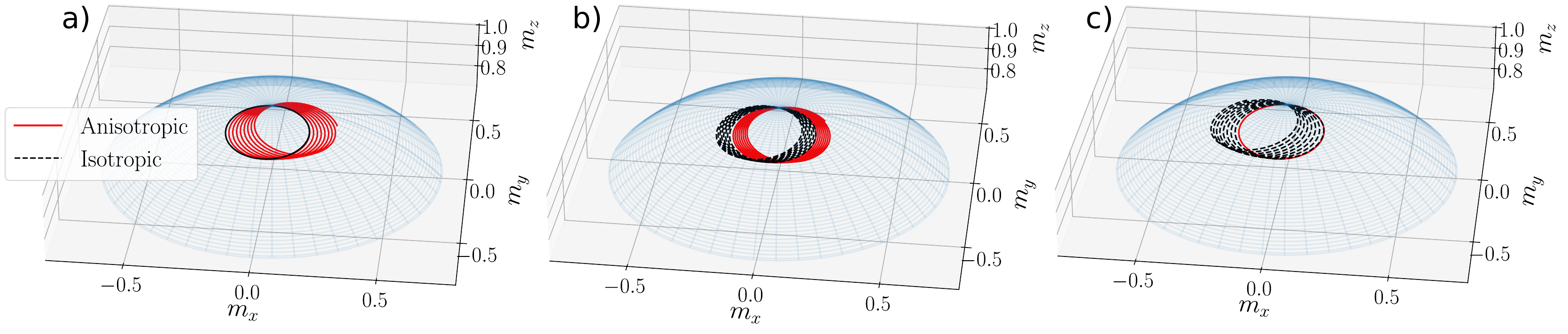}
    \caption{\small Path of the magnetization $\mm(t)$ in the absence of damping, but with a finite start velocity. In the case of isotropic nutation term, $\mm(t)$ precesses around a fixed nutation center, but the anisotropy due to the system being in a magnetically ordered state can make the nutation center move. The nutation term $I$ is fixed, but $\gamma H I$ ranges from 0 in panel a), 0.008 in b) and approximately 0.016 in panel c). The last panel shows how the nutation center motion can be entirely canceled by applying an appropriate external magnetic field $H$.}
    \label{fig:remove_resonance}
\end{figure*}

The fact that precession can be affected by $\vb{I}$ is already known \cite{Olive2015}, but here we also see the effect of the magnetization $m_0$, which one might expect to manifest somewhat like an addition to the effective magnetic field. The picture is not as clear, however, since we are considering the $\omega^2/m_0^2$ term here and not the zeroth-order term. For example, Eq.\ \eqref{eq:intertia_omega_then_q} shows that the nutation term is not explicitly dependent on the sign of $m_0$. Fig.\ \ref{fig:remove_resonance} illustrates an aspect of the expectation that the anisotropy manifests like a magnetic field. Panel a) shows a field- and damping-free comparison between the isotropic (no magnetization) and anisotropic cases with a start velocity  $\dot{\theta}_0 \neq 0$ from near the north pole. The isotropic case displays simple precession around some \emph{nutation center}, whose position can be controlled by $\dot{\mm}(t=0)$, i.e., by applying an initial fast magnetic pulse. This small but fast precession should be interpreted as the motion which superposes on the well-known precession caused by an external field to yield precession with nutation like in Fig.\ \ref{fig:faster_dynamics}.  The time scale of this precession is purely set by $I$, but as mentioned, its magnitude is controlled by the initial conditions. The path in the anisotropic case, which has magnetization, consists of a \emph{precessional} motion around the north pole with loops. This can be viewed as the same precession seen in the isotropic case, but now with a nutation center that moves around the pole because of the effective magnetic field. Next, we apply a magnetic field in the positive $z$ direction, which we know from Fig.\ \ref{fig:faster_dynamics} causes clockwise precession. Panel b) shows how this slows down the motion of the nutation center in the anisotropic case, and sets it in clockwise motion in the isotropic case. Finally, Fig.\ \ref{fig:remove_resonance} c) shows how one can tune the applied magnetic field to exactly cancel out the anisotropic case's nutation center motion, thereby removing this third resonance.

These results suggest that the spin dynamics will be affected differently at the ultrafast timescale, depending on whether or not the system is in a magnetically ordered state. Specifically, an upwards shift of the so-called nutation peak in the FMR spectrum can be expected when the nutation term is $\text{diag}(I,I,0)$ instead of the usual $\text{diag}(I,I,I)$. Therefore, a magnetization $m_0$ might be a confounding variable in an experimental measurement of $I$ if $m_0$ is not taken into account since having an anisotropic $\vb{I}$ is not exactly the same as having a reduced isotropic $I$ due to the extra term and the modified $\ddot{\theta}$ coefficient in the $\ddot{\theta}$ equation. Additionally, Fig.\ \ref{fig:remove_resonance} demonstrates that a \emph{third} frequency scale can be present in the system, giving rise to yet another peak in an FMR spectrum. We have demonstrated that such a resonance can be removed by applying a suitably large magnetic field to arrest the motion of the nutation center. We propose this more detailed prediction of a three-peak spectrum as a fingerprint for the presence of nutation in conducting ferromagnets and normal metal-ferromagnet heterostructures with weak Gilbert damping.

\section{Conclusions}
In this paper, we have extended previous calculations of the RKKY interaction in a two-dimensional metallic ferromagnet to the case of \emph{finite} magnetization stabilized by easy-axis anisotropy, relevant for normal metal-ferromagnet heterostructures with possible future spintronics applications. With these applications in mind, we also show a straightforward method to compute the full frequency dependence of the magnetization dynamics in such heterostructures, with special emphasis on the nutation term in the Landau-Lifshitz-Gilbert equation. This term, $\mm\cross(\vb{I}\cdot\ddot{\mm})$, appears as the next correction to the magnetization dynamics after the well-known Gilbert damping term $\mm \cross \dot{\mm}$. For sufficiently large magnetization, the nutation term becomes anisotropic, $\vb{I}=\text{diag}(I,I,0)$, which changes the ferromagnetic resonance spectrum in two ways: it shifts the nutation peak upwards in frequency, and another peak can appear depending on the strength of the external magnetic field.
\ \\

\section*{Acknowledgments}
We acknowledge helpful discussions with A. Brataas and A. Qaiumzadeh. This work was supported by the Research Council of Norway (RCN) through its Centres of Excellence funding scheme, Project No. 262633, ``QuSpin'', RCN Project No. 323766, as well as as well as COST Action CA21144  ``Superconducting Nanodevices and Quantum Materials for Coherent Manipulation".
\bibliography{main.bib}

\appendix
\section{Integrating Out The Fermions}
\label{sec:integrate_out_fermions}
To integrate out the fermions, we perform the integral
\begin{equation}
\label{eq:gaussian}
    I_\mathrm{NM+I}[\vb{S}] = \int \mathcal{D} \psi \mathcal{D} \psi^\dagger \e{-S_\mathrm{NM} [\psi, \psi^\dagger] - S_\mathrm{int} [\psi, \psi^\dagger, \vb{S}]},
\end{equation}
where, suppressing their dependence on $\psi, \psi^\dagger, \vb{S}$, the actions are given by
\begin{align}
  S_\text{NM} + S_\text{int} &= \int_0^\beta \dd \tau \int\dd[2]{r} \psi^\dagger(\rr, \tau)  \left[G^{-1}(\rr, \partial_\tau)\right. \\ \nonumber
  &\phantom{QQQQQQQQQQ} \left. \vphantom{G^{-1}}+ B(\rr, \tau, \vb{S}) \right] \psi(\rr, \tau) 
\end{align}
with the definitions
\begin{align}
    G^{-1}(\rr, \partial_\tau) &= \partial_\tau - \mu -m_0\sigma_z - \frac{\hbar^2}{2m} \boldsymbol{\grad}^2 \\
    B(\vb{S}) &= -2\bar{J} \boldsymbol{\sigma}\cdot \vb{S}(\rr, \tau).
\end{align}
We will employ the Fourier transform conventions
\begin{align}
    &\psi(\rr,\tau) = \frac{1}{\beta} \sum_{n} \int \frac{\dd[2]{k}}{(2\pi)^2} \psi_{n}(\kk) \e{i(\kk\cdot \rr - \omega_n \tau)} \label{eq:FT_psi_general_forward}\\
    & \psi_{n}(\kk) = \int_0^\beta \dd{ \tau} \int \dd[2]{r} \psi(\rr, \tau) \e{-i(\kk\cdot\rr  - \omega_n \tau)},\label{eq:FT_psi_general_backward}
\end{align}
where $\rr, \kk$ are dimensionless variables scaled by the lattice constant $a$, which we take to be $a=1$. The Fourier transformed version of the actions, then, is
\begin{align}\nonumber
    &S_\text{NM} + S_\text{int} = \int\frac{\dd[2]{k_1}}{(2\pi)^2}\frac{\dd[2]{k_2}}{(2\pi)^2}\sum_{n_1, n_2} \psi_{k_1, n_1}^\dagger \Bigl ( G^{-1}_{\kk_1, \kk_2; n_1, n_2} \\
    &\phantom{QQQQQQQQQQQQQQ}+ B_{\kk_1, \kk_2; n_1, n_2} \Bigr ) \psi_{k_2, n_2},
\end{align}
now with
\begin{align}
    G^{-1}_{\kk_1, \kk_2; n_1, n_2} &= \frac{(2\pi)^2 \delta(\kk_1- \kk_2)\delta_{n_1, n_2}}{\beta}  \bigl ( d_0(k_1) - m_0\sigma_z \bigr) \\
    B_{\kk_1, \kk_2; n_1, n_2} &= \frac{- 2\bar{J}}{\beta^{2}} \boldsymbol{\sigma} \cdot \vb{S}(\kk_1 - \kk_2, i\omega_{n_1} - i\omega_{n_2}),
\end{align}
and
\begin{align}
    d_0(k) &= - i\omega_n - \mu + \varepsilon_{\kk} \\
    \varepsilon_{\kk} &= \frac{\hbar^2}{2m}\kk^2,
\end{align}
using the shorthand $k_i = (i\omega_{n_i}, \kk_i)$. The integral
\begin{align}
    I_\mathrm{NM+I}[\vb{S}] &= \int \mathcal{D} \psi \mathcal{D} \psi^\dagger \e{-\psi^\dagger\left(G^{-1} + B\right)\psi}
\end{align}
is Gaussian, allowing us to compute the effective spin action in
\begin{align}
    Z &=\int \mathcal{D} \vb{S} \e{-S_\mathrm{eff}[\vb{S}]}  =\int \mathcal{D} \vb{S} \e{-S_\mathrm{FM}[\vb{S}]} \e{\Tr \ln \left(G^{-1} + B\right)}.
\end{align}
We compute this to second order in spin interactions by Taylor expanding the logarithm, such that
\begin{equation} \label{eq:Seff_TrABAB_appendix}
\begin{aligned}
    S_\text{eff}[\vb{S}] &\approx  S_{\text{FM}}[\vb{S}] - \Tr \ln G^{-1} - \Tr \left(GB\right) \\
    &\phantom{QQ}+ \frac{1}{2} \Tr \left( GB GB\right) ,
\end{aligned}
\end{equation}
where Tr denotes a trace over all quantum  numbers. $G$ is diagonal in spin, momentum, and frequency, so
\begin{align}
    G_{\kk_1, \kk_2; n_1, n_2} &=  \frac{\beta}{(2\pi)^2} \frac{ \bigl(d_0(k_1) \mathbb{I} + m_0\sigma_z\bigr)  \delta(\kk_1 - \kk_2) \delta_{n_1, n_2}}{\bigl( i\omega_{n_1} - E_{\kk_1+} \bigr) \bigl( i\omega_{n_1} - E_{\kk_1 -} \bigr)},
\end{align}
with the spin-split fermion band energies given by $E_{\kk\pm} = \varepsilon_\kk - \mu \pm m_0$. Therefore, using properties of the trace over Pauli matrix products,
\begin{widetext}
\begin{align}
    \Delta S_\text{eff}^{(2)}[\vb{S}] &= \frac{1}{2} \frac{4\bar{J}^2}{\beta^{2}} \int \frac{\dd[2]{k_1}}{(2\pi)^2}\int \frac{\dd[2]{k_2}}{(2\pi)^2} \sum_{n_1,n_2} \frac{\tr \bigl \{ \bigl[d_0(k_1)\mathbb{I} + m_0\sigma_z\bigr] \bsigma \cdot \vb{S}(k_1 - k_2) \bigl[d_0(k_2)\mathbb{I} + m_0\sigma_z\bigr] \bsigma \cdot \vb{S}(k_2 - k_1) \bigr \}}{(i\omega_{n_1} - E_{\kk_1 +})(i\omega_{n_1} - E_{\kk_1 -})(i\omega_{n_2} - E_{\kk_2 +})(i\omega_{n_2} - E_{\kk_2 -})} \\
    &= \frac{4\bar{J}^2}{\beta^{2}} \int \frac{\dd[2]{k_1}}{(2\pi)^2}\int \frac{\dd[2]{k_2}}{(2\pi)^2} \sum_{n_1,n_2} \frac{S_\alpha(k_1-k_2) S_\beta(k_2-k_1)}{D(k_1) D(k_2)} \bigl \{ [d_0(k_1) d_0(k_2) + m_0^2(2\delta_{\alpha z} - 1)] \delta_{\alpha \beta}  \\ \nonumber
    &\phantom{QQQQQQQQQQQQQQQQQQQQQQQQ} + i  m_0 \epsilon_{z \alpha \beta} [d_0(k_2) - d_0(k_1)] \bigl\},
\end{align}
\end{widetext}
This has a term which is off-diagonal in spin, but its integrand is odd under the exchange $k_1 \leftrightarrow -k_2$, so it integrates to zero. The diagonal term -- the induced exchange interaction -- is nonzero, however, such that what remains is the expression in Eq.\ \eqref{eq:S_eff_after_spin_trace} in the main text.

\section{Real-Time Spin-Spin Interactions}
\label{sec:to_real_time}
Defining $i\omega_\nu = i\omega_{n_1} - i\omega_{n_2}$ and $i\omega_n = i\omega_{n_1}$, one can perform a partial fraction decomposition in $i\omega_n$ on 
\begin{widetext}
\begin{align}
    \frac{-(i\omega_n + \mu - \varepsilon_{\kk_1})(i\omega_n - i\omega_\nu + \mu - \varepsilon_{\kk_2}) - M_\alpha}{(i\omega_n - E_{\kk_1 +})(i\omega_n - E_{\kk_1 -})(i\omega_n - i\omega_\nu - E_{\kk_2 +})(i\omega_n - i\omega_\nu - E_{\kk_2 -})},
\end{align}
which enters in Eq.\ \eqref{eq:S_eff_after_spin_trace}. Summing over $n$ in each of the resulting terms of the form
\begin{align}
    \frac{1}{\beta} \sum_n \frac{1}{i\omega_n - E} = f(E),
\end{align}
where $f(E)$ is the Fermi-Dirac distribution, leads to an action of the form
\begin{align}
     \Delta S_\text{eff}^{(2)}[\vb{S}] &= \frac{4\bar{J}^2}{\beta}\int \frac{\dd[2]{k_1}}{(2\pi)^2}\int \frac{\dd[2]{k_2}}{(2\pi)^2} \sum_{i\omega_\nu} S_\alpha(\kk_1 - \kk_2, i\omega_\nu) S_\alpha(\kk_2-\kk_1, -i\omega_\nu) \\ \nonumber
    &\phantom{QQQQQQQQQQQQQ} \times \sum_{s_1, s_2 = \pm} \frac{f(E_{\kk_1 s_1}) - f(E_{\kk_2 s_2})}{4} \,\frac{ 1 + s_1 s_2 (2\delta_{\alpha z} - 1)}{i\omega_\nu - (E_{\kk_1 s_1} - E_{\kk_2 s_2})}.
\end{align}
 To perform the analytic continuation to real frequencies and then real time, we convert the $i\omega_\nu$ sum to an integral by letting $\beta \to \infty$, such that $\sum_{i\omega_\nu} \to \beta\int\dd{\omega}/2\pi$. To obtain the real-time action, we substitute $i\omega \to \omega + i \delta$, such that the action is read off of the quantity $\int \mathcal{D}\vb{S}\,\exp (iS_{\text{eff}}^{\text{(RT)}})$. Expressed in terms of the shifted momenta $\bq = \kk_1 - \kk_2$ and $\kk = \kk_2$, the two-spin part of the action is then
\begin{align} 
    \Delta S^{(\text{2, RT})}_{\text{eff}} &= \frac{-4\bar{J}^2}{2\pi} \int \frac{\dd[2]{q}}{(2\pi)^2} \int \dd{\omega} S_\alpha(\bq, \omega + i\delta) S_\alpha(-\bq, -\omega - i\delta) \chi_\alpha(\bq, \omega),
\end{align}
where
\begin{align}\label{eq:chi_alpha_definition}
    \chi_\alpha(\bq, \omega) &= \frac{1}{4} \int \frac{\dd[2]{k}}{(2\pi)^2} \sum_{s_1, s_2=\pm} \bigl[ f(E_{\kk+\bq, s_1}) - f(E_{\kk s_2}) \bigr] \frac{1+s_1 s_2(2\delta_{\alpha z} - 1)}{\omega - \bigl [\varepsilon_{\kk+\bq} - \varepsilon_{\kk} + (s_1-s_2)m_0 \bigr] + i\delta},
\end{align}
as stated in Eq.\ \eqref{eq:chi_alpha_start} in the main text.

The static RKKY term and the nutation term are obtained from the real part of $\chi_\alpha(\bq, \omega)$, as mentioned in Section \ref{sec:omega_to_time}. Under the integral sign, we now employ the identity
\begin{align}
    \frac{1}{X + i\delta}  = \mathcal{P} \frac{1}{X} - i \pi \sgn(\delta)\delta^{(1)}(X),
\end{align}
where $X = \omega - \bigl [\varepsilon_{\kk+\bq} - \varepsilon_{\kk} + (s_1-s_2)m_0\bigr] \in \mathbb{R}$ and $\delta^{(1)}(X)$ is the one-dimensional Dirac delta function. Thus, the principal part of the $\kk$ integral at $\delta = 0$ is $\Re \chi_\alpha(\bq, \omega)$. Shifting $\kk \to -\kk - \bq$ in the $f(E_{\kk+\bq, s_1})$ term and switching $s_1 \leftrightarrow s_2$ in the $f(E_{\kk s_2})$ term allows us to write
    \begin{align}
        &\Re \chi_\alpha(\bq, \omega) = \frac{1 }{4 (2\pi)^2} \sum_{s_1, s_2} [1+s_1 s_2(2\delta_{\alpha z}-1)] \frac{\kF^2}{\mu} \int_0^{\kFs{s_1}}\frac{\dd{k}}{\kF} \frac{k}{\kF}  \mathcal{P}\int_0^{2\pi} \dd{\phi}\\\nonumber
        &\phantom{QQ} \times \left \{ \frac{1}{\nu + \bigl [ \tilde{q}^2 + 2\tilde{k}\tilde{q}\cos\phi  - (s_1-s_2)m_0/\mu \bigr]} - \frac{1}{\nu - \bigl [ \tilde{q}^2 + 2\tilde{k}\tilde{q}\cos\phi  - (s_1-s_2)m_0/\mu \bigr]}\right\} \\ \label{eq:chi_before_Phi}
        &=\frac{\kF^2}{4\mu (2\pi)^2} \sum_{s_1, s_2} [1+s_1 s_2(2\delta_{\alpha z}-1)]  \int_0^{\frac{\kFs{s_1}}{\kF}}\dd{\tilde{k}} \frac{\tilde{k}}{2\tilde{k}\tilde{q}} \mathcal{P}\int_0^{2\pi} \dd{\phi}  \left ( \frac{1}{\nu_+/2\tilde{k} + \cos\phi} - \frac{1}{ \nu_-/2\tilde{k} - \cos\phi }\right),
    \end{align}
where we employ the notation $\tilde{q}=q/\kF$, $\tilde{k}=k/\kF$, and $\nu = \omega / \mu$. The polar coordinate $\phi$ above is the angle between $\bq$ and $\kk$, and only for some values of
\begin{align}
    \nu_\pm &= \frac{\nu \pm \left[\tilde{q}^2 - (s_1 - s_2)\frac{m_0}{\mu}\right]}{\tilde{q}}
\end{align}
does the integrand have singularities at certain $\phi$ values. Note that this Appendix uses a different notation for $\nu, \nu_\pm$ than the main text. The angular integrals are of the form \cite{Mihaila2011}
\begin{align}
    \Phi(a) = \mathcal{P} \int_0^{2\pi} \dd{\phi} \frac{1}{a + \cos \phi} = \frac{2\pi}{\sqrt{a^2-1}} \theta(\abs{a} -1)\sgn a,
\end{align}
where we arrived at the expression on the right-hand side by substituting in a complex variable, $z = \e{i\phi}$, on the unit circle and applying the residue theorem. Inserting $\Phi\left(\pm \nu_\pm/2\tilde{k}\right)$ in Eq.\ \eqref{eq:chi_before_Phi} yields
\begin{align}
    &\Re \chi_\alpha(\bq, \omega) =\frac{2m}{\hbar^2} \frac{1}{8\pi\tilde{q}} \sum_{s_1, s_2} [1+s_1 s_2(2\delta_{\alpha z}-1)] \int_0^{\frac{\kFs{s_1}}{\kF}}\dd{\tilde{k}} \tilde{k} \left[ \frac{\sgn(\nu_+)\theta \left( \abs{\tilde{\nu_+}}- 2\tilde{k} \right)}{\sqrt{\left(\nu_+\right)^2 - 4\tilde{k}^2}} - \frac{\sgn(\nu_-)\theta \left( \abs{\tilde{\nu_-}} - 2\tilde{k} \right)}{\sqrt{\left( \nu_- \right)^2 - 4\tilde{k}^2}} \right].
\end{align}
When $\abs{\nu_\pm} > 2\kFs{s_1}/\kF$, the step function is always 1, so in that case,
\begin{align}
    \int_0^{\kFs{s_1}/\kF}\dd{\tilde{k}} \tilde{k} \frac{\sgn(\nu_\pm)\theta \left( \abs{\tilde{\nu_\pm}}- 2\tilde{k} \right)}{\sqrt{\left(\nu_\pm\right)^2 - 4\tilde{k}^2}} &= \frac{\sgn(\nu_\pm)}{4} \left [ \abs{\nu_\pm} - \sqrt{\nu_\pm^2 - \left (\frac{2\kFs{s_1}}{\kF}\right)^2} \right],
\end{align}
and when $\abs{\nu_\pm} < 2\kFs{s_1}/\kF$ the upper integral limit becomes $\abs{\nu_\pm}/2$ such that
\begin{align}
    \int_0^{\kFs{s_1}/\kF}\dd{\tilde{k}} \tilde{k} \frac{\sgn(\nu_\pm)\theta \left( \abs{\tilde{\nu_\pm}}- 2\tilde{k} \right)}{\sqrt{\left(\nu_\pm\right)^2 - 4\tilde{k}^2}} &=  \frac{\sgn(\nu_\pm)}{4} \abs{\nu_\pm} = \frac{\nu_\pm}{4}.
\end{align}
The $\nu_+ - \nu_-$ term is present for all $q, \nu$, so
\begin{align}
    &\Re \chi_\alpha(\bq, \omega) =\frac{2m}{\hbar^2} \frac{1}{32\pi\tilde{q}} \sum_{s_1, s_2} [1+s_1 s_2(2\delta_{\alpha z}-1)] \left [ \nu_+ - \nu_-  - \sgn(\nu_+) \theta\left( \abs{\nu_+} - \frac{2\kFs{s_1}}{\kF} \right)\sqrt{\nu_+^2 - \left(\frac{2\kFs{s_1}}{\kF}\right)^2}\right. \\\nonumber
    &\phantom{QQQQQQQQQQQQQQQQQQQQQQQQQQQQQQQqq}\left. + \sgn(\nu_-) \theta\left( \abs{\nu_-} - \frac{2\kFs{s_1}}{\kF} \right)\sqrt{\nu_-^2 - \left(\frac{2\kFs{s_1}}{\kF}\right)^2}\right].
\end{align}
To arrive at Eq.\ \eqref{eq:full_chi}, we insert $\alpha = z$ or $\alpha =x,y$ and sum over $s_2$. 

\section{Static RKKY interaction for $m_0\neq 0$}
\label{sec:aristov}
In this Appendix, we extend the results of 
Ref. \cite{Aristov1997}
to the case of non-zero magnetization, $m_0 \neq 0$. The starting point for this calculation, as mentioned in the main text, is
\begin{align}
    &\chi_\alpha(\rr_1 - \rr_2, i\omega_\nu) = \frac{1}{\beta}\sum_n \iint \frac{\dd[2]{k_1}}{(2\pi)^2} \frac{\dd[2]{k_2}}{(2\pi)^2} \frac{\left[m_0^2(2\delta_{\alpha z} - 1) +(i\omega_n + \mu - \varepsilon_{\kk_1})(i\omega_n - i\omega_{\nu} + \mu - \varepsilon_{\kk_2})\right]\e{i(\kk_1 - \kk_2)\cdot(\rr_2 - \rr_1)}}{(i\omega_n - E_{\kk_1+})(i\omega_n - E_{\kk_1-})(i\omega_n - i\omega_\nu - E_{\kk_2 +})(i\omega_n - i\omega_\nu - E_{\kk_2 -})},
\end{align}
which can be written as a sum of terms, each of which is separable in $\kk_1,\kk_2$. Utilizing this fact, one can perform a partial fraction decomposition in $i\omega_n$ on each of these factors which depend on only one momentum. Some straightforward rearrangements lead to the expression for $\chi_\alpha(\Delta \rr, i\omega_\nu = 0)$ given in the main text, Eq.\ \eqref{eq:chi_aristov_comp_GG}. What sets our calculation apart from that of Ref. \cite{Aristov1997} is the addition of the term $\mp m_0$ in
\begin{align}
    G_\pm(i\omega_n, \Delta \rr) &= \int\frac{\dd[2]{k}}{(2\pi)^2} \frac{\e{i\kk\cdot \Delta \rr}}{i\omega_n - \varepsilon_{\kk} + \mu \mp m_0}.
\end{align}
Referring to Ref.\ \cite{Aristov1997} for some of the technical details, we find
\begin{align}
    G_\pm(i\omega_n, \Delta \rr) &= -\frac{1}{2\pi} \frac{2m}{\hbar^2} K_0\left(\sqrt{-\frac{2m}{\hbar^2} z_\pm \Delta r^2}\right)
\end{align}
by replacing the chemical potential $\mu$ of Ref. \cite{Aristov1997}  with $\mu \to \mu \mp m_0$ and $m \to m/\hbar^2$. Here, $z_\pm = i\omega_n + \mu \mp m_0$, and as noted in Ref \cite{Aristov1997}, the branch of the square root in the argument of the modified Bessel function of the second kind $K_0$ must be chosen such that it has positive real part. $G_\pm$ therefore has a branch cut along the real axis when viewed as a function of $z_\pm$, which will become relevant momentarily. We first compute 
\begin{align}
    \chi_z(\Delta \rr, i\omega_\nu= 0) &= -\frac{1}{2\beta} \sum_n \sum_{s=\pm}  G_{s}(i\omega_n, \Delta \rr)G_{s}(i\omega_n, -\Delta \rr)\\
    &=-\frac{1}{2\beta} \sum_n \sum_{s=\pm}  \left[-\frac{1}{2\pi} \frac{2m}{\hbar^2} K_0\left(\sqrt{-\frac{2m}{\hbar^2} z_s \Delta r^2}\right)\right]^2
\end{align}
because the $\alpha = z$ case is where an exact analytical expression can be found with the method presented here. $\chi_{\alpha \neq z}(\Delta \rr, i\omega_\nu=0)$ can be found analytically to an excellent approximation by forming an expression analogous to the expression for $\chi_z(\Delta \rr, i\omega_\nu= 0)$. To evaluate the above sum
\begin{align}
    X_\pm(\Delta \rr) \equiv \frac{-1}{\beta} \sum_n K_0\left(\sqrt{-\frac{2m}{\hbar^2} z_\pm \Delta r^2}\right)^2,
\end{align}
we slightly modify the standard Matsubara-sum procedure in the following way. Assuming $\mu > m_0$, we take 
\begin{align}
    f_\pm(z) = \frac{-\beta}{1 + \e{\beta(z - \mu \pm m_0)}}
\end{align}
as the counting function used with the residue theorem to recast the sum as the contour integral
\begin{align}
    X_\pm(\Delta \rr) = \frac{-1}{\beta} \frac{1}{2\pi i} \lim_{\Omega \to \infty} \oint_{Q_1(\Omega), Q_4(\Omega)} \dd{z} f_\pm(z) K_0\left(\sqrt{-\frac{2m}{\hbar^2} z \Delta r^2}\right)^2
\end{align}
going counterclockwise around each of the quarter circles $Q_1(\Omega), Q_4(\Omega)$ with radius $\Omega$ in the first and fourth quadrant of the complex plane, both infinitesimally near the real axis and their respective part of the imaginary axis. Had there been no branch cut in $K_0$, the contours $Q_1$ and $Q_4$ could have been merged to a half circle in the right half plane, thus making $X_\pm(\Delta \rr) =0$. The cut, however, is present, so
\begin{align}
    X_\pm(\Delta \rr) &= \frac{-1}{2\pi i} \left [ \int_0^{\infty} \dd{\epsilon} \frac{f_\pm(\epsilon + i\delta)}{\beta} K_0\left ( \sqrt{-\frac{2m}{\hbar^2}(\epsilon+i\delta) \Delta r^2} \right)^2  + \int_\infty^0 \dd{\epsilon} \frac{f_\pm(\epsilon - i\delta)}{\beta} K_0\left ( \sqrt{-\frac{2m}{\hbar^2}(\epsilon - i\delta) \Delta r^2} \right)^2  \right]
\end{align}
because the integrals along the imaginary axis cancel and the integrand is exponentially suppressed along the circular arcs of $Q_1$ and $Q_4$ on account of $\Re z$ being strictly positive. We take the $\delta \to 0$ limit first and use 
\begin{align}
    K_0\left ( \sqrt{-\frac{2m}{\hbar^2}(\epsilon+i\delta) \Delta r^2} \right) &= \frac{\pi i}{2}H_0^{(1)}\left ( \sqrt{\frac{2m}{\hbar^2}\epsilon \Delta r^2} \right)\\
    K_0\left ( \sqrt{-\frac{2m}{\hbar^2}(\epsilon-i\delta) \Delta r^2} \right) &= \frac{-\pi i}{2}H_0^{(2)}\left ( \sqrt{\frac{2m}{\hbar^2}\epsilon \Delta r^2} \right),
\end{align}
where the Hankel functions can be rewritten in terms of the Bessel functions of the first and second kind as $H_0^{(1)}(z) = J_0(z) + iY_0(z)$ and $H_0^{(2)}(z) = J_0(z) - iY_0(z)$. Next, taking the zero-temperature limit $\beta\to\infty$, the factors $-f_\pm(\epsilon)/\beta$ become step functions such that
\begin{align}
    X_\pm(\Delta \rr) &=  \frac{-\pi}{2} \int_0^{\mu \mp m_0} \dd{\epsilon} J_0\left ( \sqrt{2\epsilon m(\Delta r)^2} \right) Y_0\left ( \sqrt{2\epsilon m(\Delta r)^2} \right) \\
    &= \frac{-\pi}{2} (\mu \mp m_0) \left [ J_0(\kF_\pm \Delta r) Y_0(\kF_\pm \Delta r) + J_1(\kF_\pm \Delta r) Y_1(\kF_\pm \Delta r) \right].
\end{align}
Inserting this result into 
\begin{align}
    \chi_z(\Delta \rr, i\omega_\nu = 0) = \frac{1}{8\pi^2} \left(\frac{2m}{\hbar^2}\right)^2 \bigl ( X_+(\Delta \rr) + X_-(\Delta \rr) \bigr)
\end{align}
then yields Eq.\ \eqref{chi_z-static}.

\section{Landau-Lifshitz-Gilbert Equation with Anisotropy}
To arrive at Eqs. \eqref{eq:llg_theta} and \eqref{eq:llg_phi}, valid for $\vb{H}=H\hat{\vb{z}}$, we employ the polar angle $\theta$ and azimuth angle $\phi$, which allows us to rewrite the magnetic-field term in 
\begin{align}
    \dot{\mm} &=  \mm \cross \left [ \gamma\vb{H} + \alpha \dot{\mm} + \left ( \vb{I} \cdot \ddot{\mm} \right)\right]
\end{align}
as $\gamma \mm \cross \vb{H} = \gamma H \left ( \hat{\vb{x}}\sin\theta \sin\phi - \hat{\vb{y}}\sin\theta \cos\phi \right)$. The time derivative of $\mm$ is $\dot{\mm} = \hat{\vb{x}}(\dot{\theta} \cos \theta \cos \phi - \dot{\phi} \sin \theta \sin\phi) + \hat{\vb{y}}(\dot{\theta} \cos\theta \sin\phi + \dot{\phi} \sin\theta \cos\phi) - \hat{\vb{z}}\dot{\theta} \sin \theta$, so
\begin{align}
    \alpha \mm \cross \dot{\mm} = \alpha\begin{pmatrix}
         -\dot{\theta}\sin\phi - \dot{\phi}\sin\theta \cos\theta \cos\phi \\
         \dot{\theta} \cos\phi - \dot{\phi} \sin\theta \cos\theta \sin \phi\\
         \dot{\phi}\sin^2 \theta
    \end{pmatrix}.
\end{align}
The expression for the next term, containing $\ddot{\mm}$, is rather large, but it can be simplified by looking at its $z$ component first. This turns out to be $[\mm\cross\left ( \vb{I} \cdot \ddot{\mm} \right)]_z = I\ddot{\phi}\sin^2\theta + 2I \dot{\theta}\dot{\phi}\sin\theta \cos\theta$, which lets us write down the entire $z$ component of the LLG equation. It reads
\begin{align}
    -\dot{\theta} \sin \theta &= \alpha \dot{\phi} \sin^2\theta + I \ddot{\phi} \sin^2\theta + 2I \dot{\theta} \dot{\phi} \sin \theta \cos \theta,
\end{align}
from which Eq.\ \eqref{eq:llg_phi} for $\ddot{\phi}\sin\theta$ immediately follows. Replacing every $\ddot{\phi}\sin\theta$ in $\mm\cross\left ( \vb{I} \cdot \ddot{\mm} \right)$ by the expression in Eq.\ \eqref{eq:llg_phi} leads to several cancellations, such that the $x$ and $y$ components of the LLG equation read
\begin{align}
    \begin{pmatrix}
        -\dot{\phi} \sin\theta \sin\phi\\
        \dot{\phi} \sin\theta \cos\phi
    \end{pmatrix} &= \gamma H \begin{pmatrix}
        \sin\theta \sin\phi\\
        -\sin\theta \cos\phi
    \end{pmatrix} + \alpha \begin{pmatrix}
        -\dot{\theta} \sin\phi \\
        \dot{\theta} \cos\phi
    \end{pmatrix} \\\nonumber
    &\phantom{Q} +  \begin{pmatrix}
        -\ddot{\theta}\sin\phi \left(I\cos^2\theta + \tilde{I}\sin^2\theta \right) + \sin\theta\cos\theta \left[ \left(I-\tilde{I}\right) \dot{\theta}^2 \sin\phi + I\dot{\phi}^2 \sin \phi \right] \\
        \phantom{-}\ddot{\theta}\cos\phi \left(I\cos^2\theta + \tilde{I}\sin^2\theta \right) - \sin\theta \cos\theta \left[ \left(I-\tilde{I}\right) \dot{\theta}^2  \cos\phi + I\dot{\phi}^2  \cos \phi \right]
    \end{pmatrix}.
\end{align}
We now multiply the top equation by $\sin \phi$ and the bottom equation by $\cos \phi$ and then subtract the bottom from the top equation, to obtain
\begin{align}
    -\dot{\phi} \sin \theta &= \gamma H \sin\theta - \alpha \dot{\theta} - \ddot{\theta} \left(I\cos^2\theta + \tilde{I}\sin^2\theta \right) + I \dot{\phi}^2 \sin\theta \cos\theta + \left(I-\tilde{I}\right) \dot{\theta}^2 \sin\theta \cos\theta.
\end{align}
Eq.\ \eqref{eq:llg_theta} immediately follows from the above equation.

\end{widetext}
\end{document}